# Developing a Trusted Human-AI Network for Humanitarian Benefit

**Authors**: S. Kate Devitt[1,2], Jason Scholz[3], Timo Schless[4], Larry Lewis[5]

[1]Trusted Autonomous Systems, Australia

[2] University of Queensland, Australia

[3] RMIT, Australia

[4] Whiteflag Foundation, Netherlands

[5] Center for Naval Analysis, Arlington, Virginia, USA

**Corresponding Author:** Dr Kate Devitt Chief Scientist of the Trusted Autonomous Systems Cooperative Research Centre in Australia and Adjunct Research Fellow at the University of Queensland. Email: k.devitt@uq.edu.au   https://orcid.org/0000-0002-6075-4969

**Co-Authors:**

Dr Jason Scholz is an Innovation Professor at the RMIT University Australia. Email: jason.scholz@rmit.edu.au   https://orcid.org/0000-0002-3656-046X

Mr Timo Schless MSc is one of the developers of the Whiteflag Protocol, and Advisor and Developer for the Whiteflag Foundation. Email: tschless@acm.org  https://orcid.org/0000-0003-2501-6300

Dr Larry Lewis is Research Director at CNA, a research organization in Arlington, VA. Email: LEWISL@cna.org  https://orcid.org/0000-0002-2264-980X

**Abstract**: Artificial intelligences (AI) will increasingly participate digitally and physically in conflicts yet there is a lack of trusted communications with humans for humanitarian purposes. For example, in disasters and conflicts messaging and social media are used to share information, however, international humanitarian relief organisations treat this information as unverifiable and untrustworthy. Furthermore, current AI implementations can be brittle, with a narrow scope of application and wide scope of ethical risks. Meanwhile, human error can cause significant civilian harms even by combatants committed to compliance with international humanitarian law. AI offers an opportunity to help reduce the tragedy of war and better deliver humanitarian aid to those who need it. However, to be successful, these systems must be trusted by humans and their information systems, overcoming flawed information flows in conflict and disaster zones that continue to be marked by intermittent communications, poor situation awareness, mistrust and human errors. In this paper, we consider the integration of a communications protocol (the 'Whiteflag protocol'), distributed ledger 'blockchain' technology, and information fusion with artificial intelligence (AI), to improve conflict communications called "Protected Assurance Understanding Situation & Entities" PAUSE. Such a trusted human-AI communication network could provide accountable information exchange regarding protected entities, critical infrastructure, humanitarian signals and status updates for humans and machines in conflicts. Trust-based information fusion provides resource-efficient use of diverse data sources to increase the reliability of reports. AI can catch human mistakes and complement human decision making, while human judgment can direct and override AI recommendations. We examine several realistic potential case studies for the integration of these technologies into a trusted human-AI network for humanitarian benefit including mapping a conflict zone with civilians and combatants in real time, preparation to avoid incidents and using the network to manage misinformation. We finish with a real-world example of a PAUSE-like network, the Human Security Information System (HSIS), being developed by USAID, that uses blockchain technology to provide a secure means to better understand the civilian environment.

**Highlights**

- A trusted human-AI communications network for humanitarian benefit can help reduce and manage the fog of war

- Human error causes significant civilian harms even by combatants committed to complying with international humanitarian law

- Humans in disasters and conflict use messaging and social media to share information, but international humanitarian relief organisations treat this information as unverifiable and untrustworthy

- Trust in humans and machines is composed of multiple factors but can be improved through information fusion that incorporates diverse information sources, a communications protocol and use of distributed ledger technologies such as Blockchain

- Protective AI used within a trusted network, can identify protected objects and persons and assist human decision-making in conflicts to reduce civilian harms.

# 1 Introduction

21[st] century battlefields are congested and mediated. Human perception in conflict is influenced by digital information and algorithmic intervention, creating epistemic distance, increasing decision uncertainty and risk to civilians (Holland, 2021). A trusted international communications network is needed to create a better digital humanitarian response for the most vulnerable (Burleigh & Birrane, 2011; Chernobrov, 2018)[1]. Such a network is needed to improve the culture of perception, reflection, justification, and decision making in war. The UN Secretary-General explicitly proposed the requirement for real-time information exchange and coordination to protect medical services in his recommendations under Security Council resolution 2286, calling for measures such as: "Recording and mapping the presence of personnel exclusively engaged in medical duties, their means of transport and equipment, as well as hospitals and other medical facilities, and regularly updating this information, including through enhanced information exchanges and real-time coordination with medical and humanitarian actors on the ground and the use of appropriate technology (United Nations Security Council, 2016). This article provides boundary considerations (Spee & Jarzabkowski, 2009) on a proposed trusted communications network we call, 'Protected Assurance Understanding Situation & Entities' (PAUSE). A PAUSE network is designed to use technology to amplify the signals of protected persons to decision makers commensurate with human psychology, and increase accountability of unlawful targeting practices.

While important work has to be done questioning and critiquing the introduction of new technologies into the battlefield—including myths around their ability to improve ethical and lawful targeting (Bellanova et al., 2021; Suchman, 2020). This chapter treats the incursion of new technologies including digital communications, mosaic warfare, robotics, autonomous systems and artificial intelligence (RAS-AI) in war as a given. In 'War Transformed', Mick Ryan (2022) highlights 21[st] Century trends that must be addressed:

1. Increased speed of planning, decision-making, and action.
2. Large-scale conventional forces combined with the massed use of autonomous systems and extensive influence tools.

---

[1] A trusted communications network needs digital and physical infrastructure, trustworthy protocols and diverse messages from citizen journalists, professional journalists, NGOs, governments, and militaries within a conflict.

3. Human-machine integration. Autonomous systems will be full partners with human beings in the conduct of military missions.

Mixed initiative decision-making between humans and machines (Lambert and Scholz, 2005) will be a given. This article thus attempts to introduce a means to counter the ethical and legal risks associated with the increase diverse RAS-AI, information and network technologies into the battlefield.

## 1.1 War and Artificial intelligence

Artificial Intelligence (AI) is widely seen by States as imperative to national security (Netherlands Ministry of Foreign Affairs, 2023). As a result, AI-enabled systems will increasingly make their way onto the battlefield, both by militaries and by humanitarian groups working to aid civilians. While stated rationales for the use of AI tends to focus on improved military effectiveness and cost saving measures, AI also brings an opportunity to help reduce the tragedy of war and better deliver humanitarian aid to those who need it (Lewis & Ilachinski, 2022). However, to be successful, these systems must be trusted by humans and their information systems, overcoming flawed information flows in conflict and disaster zones that continue to be marked by intermittent communications, poor situation awareness, mistrust and human errors. In this way, AI must not be viewed as the naive techno-solution to harm in conflict, but one part of a systematic overhaul of the processes by which targeting decisions and accountability mechanisms for those decisions are made (Department of Defense, 2022; Lewis, 2021; Suchman, 2020).

This paper discusses practical steps to reduce the human costs of conflict and better protect providers of humanitarian assistance. We argue that this is achieved through the creation of a conflict communication standard for non-military and militaries within conflicts and indisputable record of communication transactions during conflicts. Shared information must be evaluated, justified and utilised to the satisfaction of individual parties in the conflict. A distributed ledger model of communication enables human-human, human-machine agent systems and machine agent-machine agent systems to build reputation and relay trusted information within a multi-tiered network of checks and balances. The result is a safety net: technology does not take away from human decisions, but offers additional information that can help avert human mistakes that invariably happen.

This paper is the first step; conceptualising and laying the groundwork of a human-AI network for humanitarian benefit. Next steps include value-sensitive design with stakeholders to civil-military communication in a conflict, implementation trials, and building technology readiness level (TRL) capability of the network in line with responsible innovation (van den Hoven, 2013).

# 2   The Trust Challenge

## 2.1   The Fog of War

In Bomber Command at the start of WWII, British airmen experienced the fog of war[2], including the cloud cover that hid the world from their understanding (Clausewitz, Howard, & Paret, 1976, Book ii, Ch.2.). Instead of modern global positioning, navigators relied on establishing pinpoints from the ground described as "groping" (Ch.3, Hastings, 1979). Wireless operators could pick up a loop bearing (Mason, 1992) from England, but a misjudged signal could turn the aircraft on a 180-degree reciprocal course and the Germans often jammed the wavelengths. Weather reports were inaccurate, blowing aircraft off course and speed. Visual confirmation of targets required flying so low that there was a high risk to crews from flak or enemy fighters. Lacking radar, communications between planes in formation was by Aldis lamps[3] that required visual line of sight (Duffie Jr, 2017). Thus, even though explicit instructions from Command in 1940 were not to drop bombs indiscriminately, random results were the outcome[4].

War creates unique conditions for uncertainty via intentional and inadvertent causes. Much of the error of war could be reduced if decision makers knew more had access to improved

---

[2] The term "fog of war" refers to the state of ignorance in conflict due to ambiguity. Carl von Clausewitz is credited with the first examination of the concept, though he did not use that precise phrase—attributed to Lonsdale Hale (e.g., Hale, 1897).

[3] An Aldis lamp is a signal lamp that allows light to deliver messages via morse code.

[4] The locations of bombs was in fact, so random, that Germany was genuinely unaware that Bomber Command was intending to attack a specific target or region. Still, missions were still flown, as it was deemed important that Britain was doing "something" (Hastings, 1979)

accuracy and analysis of information (Lewis, 2019b)[5]. Increasing the precision of weapons also increases the expectations of the civilian population that weapons will better avoid civilian causalities (Beier, 2003; Brown, 2007; Enemark, 2013; Walsh, 2015). Modern intelligence, surveillance and reconnaissance (ISR) technologies combined with precision effects have greatly increased the quality of justifications expected of decision-makers for their actions. For example, the information that guides a strike team into a compound after days or weeks of an ISR soak of the area can often justify (though not always) their actions in accordance with Commander's intent[6]. But protecting non-combatants and identifying combatants remains difficult in conventional warfare in high tempo environments where combatants operate in the same area with civilians (Lawfare, 2020).

## 2.2  Decision-making

The ambition of humanitarian organisations in conflict is to increase adherence to International Humanitarian Law (IHL) and the Laws of Armed Conflict (LOAC). While it is the ambition of humanitarian organizations to increase adherence to IHL/LOAC, ultimately states and their militaries are responsible for adherence to international humanitarian law. There are three action components required to achieve this: Awareness, Intent (underpinned by "will") and Capability (Lambert & Scholz, 2005). If decision-makers inside conflict zones, have *awareness* regarding what and where protected objects are, have the *intent* to abide by IHL and have the *capability* to follow-through, then better humanitarian outcomes are predicted. This paper focuses primarily on the challenge of awareness, acknowledging that awareness absent humanitarian intent or capability is ineffective and leads to a lack of trust.

Awareness can be broken down into multiple concepts, include knowledge and understanding, but also awareness of degrees of ignorance. Knowledge has traditionally been defined as "justified true belief" and represents the highest epistemic goal (Moser, 2005; Sosa, 2011). Understanding may include causal mechanisms, reasons, explanations and the meaning of what is observed (Miller, 2019). However, given the uncertainties of conflict, a sound goal to maintain trust might not be knowledge, but something short of knowledge, such as rationally-justified belief, or evidence-based decision-making. The awareness demanded by alternate epistemic frameworks such as Bayesian epistemology (Bovens & Hartmann, 2004) and evidentialism (Conee & Feldman, 2004) is that an agent is justified in making a decision if they act responsibly and proportionately given the (often uncertain and incomplete) evidence. Bayesian epistemology also provides a normative framework to guide evidence selection, valuing both the independence and diversity of information sources.

---

[5] More data by itself won't necessarily solve the problem if the structure or method of analysis cannot give you the insight you need (private communication with Dr. Beth Cardier).

[6] Noting that knowledge of one's target does not ensure abidance with requirements of *jus in bello* and *jus ad bellum* obligations.

Traditionally military information and communications technologies (ICT) have depended upon fairly narrow sets of vetted information regarded with high degrees of confidence. Future military and non-military ICT will likely draw on an internet of things including drone sensor feeds, high altitude platforms, satellites, social media, text messages and so forth plus social media messages and AI classification and recommendations to inform awareness and actions. If actors within a conflict broaden the data sources they draw on, it both increases their uncertainty and increases the potential of their awareness. In order to trust diverse information sources, their evidential value must be appraised and integrated appropriately within a larger operational picture.

Awareness of one's own uncertainty is a virtue associated with intellectual humility. We argue that decision-makers who acknowledge gaps in their knowledge and understanding are less likely to make foolhardy mistakes. So, whilst decision-makers might strive for knowledge, they are justified in the fog of war to make decisions when a certain threshold for evidence is met and the perceived risk of inaction is greater than the risk of action. The higher the humanitarian risk, the greater the evidential expectations in accordance with just war principles of discrimination and proportionality (Coates, 2016). As ISR technologies have improved, so has the expectation for militaries to hold fire under uncertainty (Ekelhof, 2018).

Perhaps surprisingly, militaries that embraced Internet technologies for decision-making through network-centric warfare (Cebrowski & Garstka, 1998; Eisenberg, Alderson, Kitsak, Ganin, & Linkov, 2018) have not necessarily invested in smarter methods to improve humanitarian protections. Despite the recent rise of digital, artificial intelligence (AI) and autonomous technologies, precision targeting and layered legal review processes, systemic situational incomprehension continues to result in unintentional harms and loss of life—and civilians bear the brunt of harm in conflict.

## 2.3 Human Error

The majority of casualties in conflict are civilians, with this harm compounded by reverberating effects of attacks (Nohle & Robinson, 2017; Roberts, 2010). Some of this harm is due to combatants that disregard requirements of international humanitarian law (IHL). For example, recent evidence uncovered by the New York Times reveal the intentionality of attacks by Russia and Syria on protected medical facilities (Khan, Hassan, Almukhtar, & Shorey, 2022; Triebert, Hill, Browne, Hurst, & Khavin, 2019). But significant harm to civilians can still occur with combatants committed to complying with IHL, such as contributors to the Counter-ISIS Coalition regarding operations in Iraq and Syria, or international forces operating in Afghanistan (Lewis, 2018, 2021). While those militaries conducted legal and policy reviews for every single strike, significant numbers of civilians were still harmed. Analysis of over 2000 incidents of civilian casualties revealed how this occurs in practice: while some cases were due to deliberate decisions that the military utility outweighed the cost to civilians, the vast majority of cases were due to human error. In these cases of human errors, either decision-makers missed indicators that civilians were present, or civilians were mistaken as combatants and attacked in that belief. Misidentifications were often a result of humans making judgments that a threat existed, either mis-associating intelligence with a specific location/individual or incorrectly ascribing hostile intent to observed behaviour. Such attacks often included a loss of situational awareness that could have helped inform a better engagement decision.

These human errors are also seen in attacks on medical facilities, as observed both in the US attack on a Médecins Sans Frontières (MSF) hospital in Afghanistan in 2015 and in multiple attacks on hospitals in Yemen by the Saudi-led coalition. Despite both reporting their location to military forces and displaying a red crescent sign, these hospitals were still attacked by military forces in the mistaken belief that they were military targets. Analysis of these inadvertent attacks reveals patterns of human errors both in deconfliction (since these structures were on the No Strike List) and in identification (since attacks failed to identify either the nature of medical facilities or the red crescent symbol marking the structure) of medical facilities (Lewis, 2019b).

Real world operations show that the deconfliction process is particularly challenging in cases of self-defence and dynamic targeting. For example, in Yemen, the majority of attacks on hospitals occurred due to dynamic targeting. Likewise, the US strike on an MSF hospital in 2015 was a dynamic targeting operation in defence of forces on the ground. Traditional planning and intelligence preparation of the battlefield, including consultation of the No Strike List, may not be optimised for short-notice operations in dynamic environments.

Likewise, the identification of medical facilities and other protected entities can be challenging in practice. Such structures may not be within established hospitals, instead being located in other facilities or even in tents. The practical identification measure of a red cross or crescent, originally from the Geneva Conventions of 1949, is not always sufficient for identification to stop attacks. Not only can the time of day or night and the presence of obscurants (e.g. dusty or cloudy conditions) affect observation of these symbols, but the type of sensor can also play a role. For example, a coloured marking will not necessarily be a discriminating feature for a pilot conducting an air strike using an infrared sensor, a type of sensor used by many modern militaries.

In summary, while the law is clear regarding the protected status of civilians and of medical facilities in armed conflict, humans make mistakes, and the limited tools and procedures available on today's battlefield for protection leave much room for such mistakes, with tragic results. The number of tragic attacks on medical facilities over the past few years point out the benefit of developing additional practical measures that can reduce the chances of such mistakes.

## 2.4  A Human-AI network

A tool that might be able to assist IHL abidance is the autonomous and rapid identification and classification of protected objects and civilians using AI and machine learning. However, unpredictability, challenge of explainability and bias of AI algorithms might also increase risk to civilians (International Committee of the Red Cross, 2019). Additionally, the usefulness of AI may be undermined by both deliberate tricking or "spoofing" and the challenges of keeping algorithms up-to-date in a changing environment (Brundage et al., 2018). Further challenges include the facts that telecommunications within conflicts remains volatile and intermittent; and that it is in militaries' interests to exploit the fog of war for strategic advantage by obfuscating operations and intentions while striving to better understand their operational environment.

Given then, that humans and AI systems have both strengths and weaknesses, we advocate that human strengths be applied to offset AI weaknesses, and AI strengths be applied to offset human fallibilities. Such a human-AI network would also automatically record available information for potential legal review.

# 3 Trusted Networks

## 3.1 Diverse Information

The rise of digital technologies in all aspects of life has changed information availability in conflicts and disasters, driving a new requirement to share among disparate groups and in new applications. Those experiencing disasters and seeking information about the disasters use messaging and social media, yet international humanitarian relief organisations treat bystander information as unverifiable and untrustworthy and these data sources do not impact organisational decision-making (Tapia, Bajpai, Jansen, Yen, & Giles, 2011).

Agencies are "reluctant to use social media, especially to gather unverified crowdsourced data" (Hiltz et al., 2020). Compounding this, military signaling has traditionally been secretive and bespoke to meet the needs of each Nation and to keep operations unknown to opposing forces. Likewise, historically humanitarian organisations have avoided sharing communication technologies with militaries to ensure their neutrality and hence immunity from harm. Yet, smart phones and social media are readily used for many purposes by: State and non-State actors to incite and engage in warfare (Singer & Brooking, 2018); by humanitarian groups to communicate regarding humanitarian needs and provision of aid; and by local populations to inform the world of the impacts of military actions. There is an opportunity for these groups to exchange these new sources of information to better meet humanitarian goals, but such exchanges must satisfy several conditions including how to manage diverse information sources and accountability for messages sent.

More recent research has developed methods and tools to analyse social media behaviours to help in humanitarian and disasters (Goolsby, Woodward, Pruulmann-Vengerfeldt, Hempson-Jones, & Falzon, 2019; Ogie et al., 2022) as well as to reveal malicious and deliberate influence campaigns relevant to evaluating peace time, grey zone and conflict information flows (Vanni, Kase, Karunasekara, Falzon, & Harwood, 2017; Weber & Neumann, 2020, 2021).

## 3.2  Information Fusion

Trust is a multi-faceted relationship that may be asymmetric between human-humans, humans-AI and AI-AI involving dimensions such as competence (including reliability, skills and experience) and integrity (including honesty, motivation and character) (Devitt, 2018). Trust can be measured as a factor of disposition, situation and learning (Hoff & Bashir, 2015). In some situations, a source may be trusted and in others not. Sources are also distributed, so information must be fused to improve trust. A predominantly autonomous fusion scheme is needed that makes best use of an information network characterised by:

- duplication (a diversity issue);
- reporting errors and error propagation (a competency issue);
- intentional errors (an integrity issue); and
- cost in terms of access to sources and risk to validate (efficiency and risk issues).

A scheme that observes all of these characteristics is hard to find in practice (Azzedin & Ghaleb, 2019). Most rely on high-trust protected sources (to avoid deception), or large-scale diversity to provide statistical evidence. Fewer still attempt to use as few sources as possible to maximise trust. A system that comprehensively attempts to tackle all of these issues called TIDY (Etuk, Norman, Şensoy, Bisdikian, & Srivatsa, 2013) is illustrated in Figure 1.

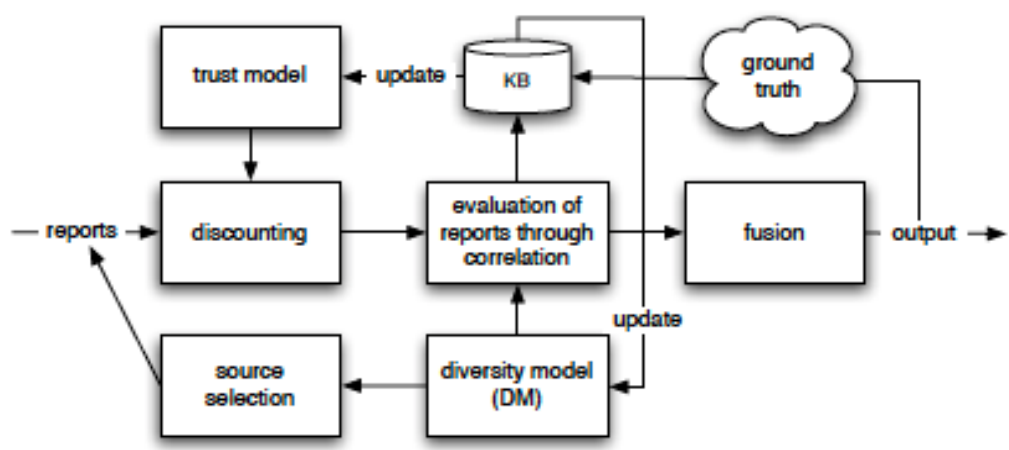

Figure 1. The framework for Trusted-based Information fusion through Diversity (TIDY).

In TIDY the trust model is based on subjective logic and enables discounting of reports to reflect the assessed reliability of the source of each report (noting all past reports are also used in this computation). The diversity model uses similarity metrics to structure the source population, so that similar sources are grouped together. Information sources with similar features are assumed likely to provide similar reports in a situation. Example features may include: organisational affiliation, known alliances, nationality, location for this report, time of this report, expertise, etc coded as numeric values. The source-selection module, uses knowledge of source diversity to sample the population of sources for evidence according to the assigned budget. Correlations of reports exploit knowledge provided in the diversity model. The knowledge base stores feedback on the fusion estimate with reference to later observed ground truth. The knowledge base also holds behavioural evidence for sources in different groups with regard to report similarity. Based on the evidence gathered, both the trust and diversity models are updated to reflect new knowledge. In the case of the diversity model, a learning process may be initiated to maintain model consistency.

The fusion approach minimises the adverse effect of large groups of unreliable sources that might collude to undermine the trustworthiness of the fusion output. Civilian-military information exchange needs to incorporate diverse information sources and fused. But this is not enough to create a trusted communications network. In addition, information must be neutral, secure and provide undeniable proof of receipt. Neutral means the network is not owned or controlled by any of its users, and its use will not compromise the impartiality of their missions. Secure means the information exchanged between military and civilian parties is authenticated, confidential and has a level of integrity commensurate with the authority from which it was sourced. For example, the integrity of information from trusted sources may be high, and from an unknown casual observer low. Proof of receipt ensures ongoing support for the course of justice, ensuring information receipt cannot be repudiated as a valid record.

## 3.3  Distributed Ledger Technology

A distributed ledger via Blockchain technology provides a mechanism for recording transactions between parties efficiently and verifiably, without the use of a centralised register (Crosby, Nachiappan, Pattanayak, Verma, & Kalyanaraman, 2016). Although theoretically not completely immune to being altered, it is practically extremely difficult with diverse and independent sources to compromise as any retrospective alteration would need to be made consistent with alteration in all down-stream blocks, and so requires a high degree of consensus[7]. Blockchain technology is provides automated, highly secure records for tracking data, transactions, contracts and even algorithm provenance across the public and private sectors allowing collaboration and integration across organisations without the need for a trusted third party, a network operator or a system owner (Efanov & Roschin, 2018; Peck, 2017). Thus blockchain is 'trust-enhancing', providing some additional trustworthiness to information exchange through reliability, but note it does not resolve other features relevant to concepts of trust such as goodwill, encapsulated interest or integrity (Devitt, 2018; Jacobs, 2020).

Blockchain lends itself to public and open networks – even if intermittent[8]. Shared information, consensus, independent validation and information security are characteristics that make blockchain of particular interest to the humanitarian sector, in order to address all sorts of practical problems related to transparency, efficiency, scale and sustainability (Coppi & Fast, 2019). Messages on a blockchain cannot disappear or be manipulated, and the origin of the information cannot be disputed. Additionally, opposing forces and neutral organisations can quickly and directly communicate to one or more other parties involved in the conflict.

## 3.4 Communications Protocol

Organisations and individuals protected under humanitarian law in conflict and disaster zones need to identify themselves in order to prevent or minimise collateral damage. A communications protocol using blockchain might contribute to a trusted civilian-military communication network, allowing combatant and neutral parties in armed conflicts to digitally communicate (see Box 1. The Whiteflag Protocol). We argue that such a protocol in combination with an information fusion scheme could be used to assess the trustworthiness of information by instant verification of the originator, authentication of reliable sources, cross-checking facts with persistent information on the blockchain to evaluate reliability of sources, confirmation by multiple sources, duress functionality, and implementation-specific measures such as filtering, blacklisting, other sources[9].

---

[7] (Efanov & Roschin, 2018; Peck, 2017).

[8] If telecommunications are intermittent, the distributed ledgers each record what information was available, where and when, to justify information available for decision retrospectively at the time, and then re-synchronises the ledger automatically when telecommunications are re-established.

---

**Box 1: The Whiteflag Protocol**

The Whiteflag protocol is a free and open standard for a digital communications protocol based on blockchain technology (Kolenbrander & Schless, 2019; "The Whiteflag Protocol," 2020). See Appendix A for the current, extensible, Whiteflag message set including protective signs, emergency signals and status signals etc... Whiteflag can be implemented for new and existing systems, (e.g. geographical information systems, command & control systems, mobile devices, beacons, transponders in cars, autonomic systems, etc.) and a variety of blockchain networks (e.g. Bitcoin and Ethereum).

Whiteflag messages may be disclosed only to trusted parties, and be hidden to others, by using encryption; they can be made available to entities with limited visibility, e.g. beyond visible range, to smart weapons, etc.; their source can be verified to establish their authenticity; they cannot be manipulated by others, and their existence is recorded permanently with undeniable proof to create transparency and help the course of justice. For example, in an extremely hostile environment, humanitarian organisations may use Whiteflag encryption to prevent information becoming available to potential hostile parties.

Whiteflag makes use of the Advanced Encryption Standard (AES) that can be set-up in multiple ways. One way is to use automatically negotiated keys for secure one-on-one communication, eliminating the need for any prior coordination. In addition, more advanced implementations are possible, where multiple sets of encryption keys are shared between participants to create different trusted subgroups. Since Whiteflag does not reveal any information on the intended recipient of the information, and the originator may choose an authentication method that only reveals its identity to trusted parties, not only the information content itself is secured, but information about the communicating parties can also be concealed.

Whiteflag has been verified and validated at Technology Readiness Level (TRL) 5 out of 9 (Jamier, Irvine, & Aucher, 2018), which means that Whiteflag technically works and the overall functionality is considered useful (Capgemini, 2018), but further operational test & evaluation activities are required to work out specific use cases, to identify and mitigate risks, and to integrate it effectively, safely and securely in new and existing systems. The Whiteflag software used for testing is open sourced and available on GitHub (Timo [ts5746]) and all test data is on the Ethereum Rinkeby Test Network (between blocks 3350000 and 6350000).

---

[9] Integration with other information resources, such as Truepic, a photo and video verification platform fighting disinformation (see https://truepic.com) has been demonstrated.

## 3.5  Protective AI

So far in the paper we have discussed the needs of decision-makers in a conflict to incorporate diverse information sources on a trusted network. It is at this juncture that we would now like to consider how AI might contribute to building a more trusted network for humanitarian benefit. A recent study of software requirements related to social media for humanitarian and emergency management applications suggested AI in half their features for priority development (see Table 5. Hiltz et al., 2020). For example, AI has been recommended for dynamically extracting information and identifying damage and severity of harms in social media images including injured, trapped or displaced people. Understandably, there are moral concerns for the potential range of uses of AI in critical roles in conflicts. While AI is likely to be incapable of a level of reasoned action sufficient to attribute moral responsibility in the near term, we argue it might today autonomously execute human value-laden decisions embedded in its design and in code. By doing so AI can perform actions to meet enhanced ethical and legal standards (Scholz & Galliott, 2018). We consider two conceptual possibilities with regards to embedded ethics within machines: MaxAI and MinAI

A maximally-just ethical machine or "MaxAI" guided by both acceptable and non-acceptable actions has the benefit of ensuring that ethically obligatory lethal action is taken, regardless of engineering foresight. That is, MaxAI is a machine that could potentially make "life and death" decisions. However, a maximally-just ethical AI requires extensive ethical engineering and may not meet the human-based judgements required under IHL. Additionally reasoning about the full scope of what is ethically permissible, including notions of proportionality and rules of engagement is a hard problem. Arguably, such an advance for machines to comprehend the human condition seems remote.

A minimally-just ethical machine or "MinAI" at the other end of the spectrum, could deal only with what is ethically impermissible. That is, MinAI could make "life" decisions. In conflict zones, these constraints are based around the need to identify and avoid "protected" objects and behaviours including lawfully-protected symbols, protected locations, basic signs of surrender (including beacons), and potentially those that are *hors de combat*. These AI problems range from easy to difficult, but not impossible, and technologies will likely continue to improve. A simple example is the ability for standard machine learning algorithms to identify symbols of protection, such as a Red Cross, in order to avert attacks. Figure 2 illustrates application of the "Faster RCNN" algorithm[10], to a mobile military hospital facility, with markings clearly and automatically identified on tents and on trucks.

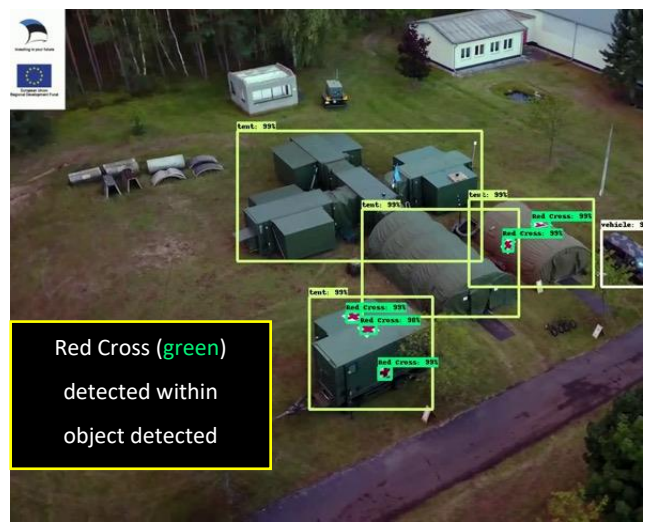

---

Figure 2. Example of open-source AI for detecting and classifying objects in a video stream including tents and vehicles, and within those object detections, the presence of protected symbols of the Red Cross.

This technology, if implemented, could potentially have saved the lives of medical workers and averted damage to medical facilities in recent Yemen security operations (Lewis, 2019a). Figure 3 illustrates examples from our own trials examining the potential for AI identification of surrender[11].

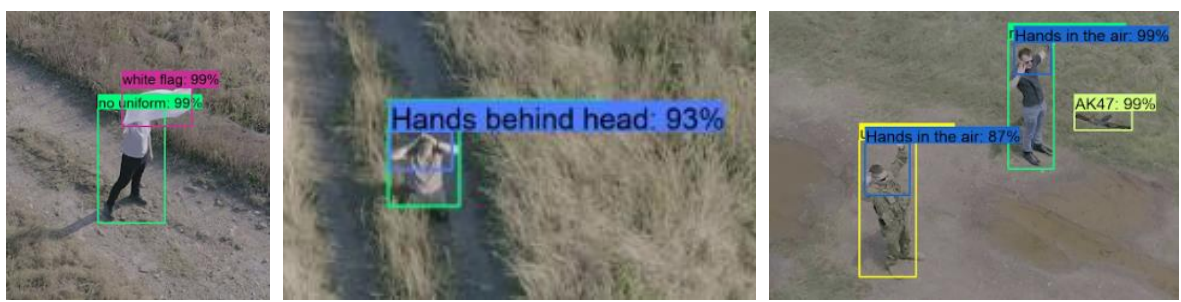

Figure 3. Highly-magnified examples of classifier detecting and classification as part of a video stream: (i) a person without uniform with a white flag, (ii) non-uniformed person with hands behind their head, and (iii) two persons each with hands in the air, which are separated from an AK47 gun.

Noting that Faster RCNN combines detection, signified by a coloured box, with classification, signified by a label. When multiple detections and classifications are made, it is possible to perform rudimentary automated reasoning. For example, in Figure 2, the yellow boxes indicate a tent, and within some of these boxes, there are other boxes indicating a symbol of the Red Cross, which implies the "Red Cross is on the tent". Protective AI also holds promise to track actions and change of state of significance, i.e. consider from the elements in Figure 3, the potential to detect surrender behaviour as a transition from a state where a person is holding a gun, then discards it, and puts their "hands in the air" separated from the "gun" on the ground.

Design of systems to achieve Protective AI based on a starting point of MinAI, must consider its own weaknesses and errors (see Appendix B. possible decision states for MinAI-enabled weapon systems). The humanitarian value added by Protective AI is when the machine correctly perceives protected objects and the human does not. From the humanitarian perspective this is essentially the addition of a safety net: it doesn't detract from anything that a human would do, and if a human makes a mistake, this is a chance for a machine-enabled capability to catch and correct that mistake. From the military perspective this is balanced against when the Protective AI incorrectly perceives an object as protected when it is legally afforded no protection, and a strike is inappropriately called off. The latter may occur as a matter of natural error and should be infrequent, however, if deliberately-caused through adversarial action this should be a matter for the creation of new legal restrictions related to perfidy.

---

[11] Work undertaken by Cyborg Dynamics Engineering and Skyborne Technologies as part of a project for the Trusted Autonomous Systems Defence Cooperative Research Centre, see https://tasdcrc.com.au/

Any Protective AI needs an appropriate legal and policy framework to inform development and use. Some policies specific to AI and data protection might include:

- Data used for machine learning object recognition, including object labelling shall be protected from tampering;
- It shall not be permitted to train neural networks starting with weights derived from unverified data sources (due to the potential to have embedded adversarial examples);
- Any delivered machine-learning system shall include not only the algorithm, but all data used in training (to ensure the executable code can be reconstructed);
- Any delivered machine-learning system shall be reproducible and repeatable entirely from its algorithm and all its training data. Etc.

# 4   A Trusted Human-AI Network

Taking now the elements of human decision-makers, a communications network and AI, we propose a trusted human-AI network that adopts a civil-military communications protocol with diverse information fusion on a distributed ledger as illustrated in Figure 4. We call this network: Protected Assurance Understanding Situation and Entities (PAUSE). The PAUSE architecture mirrors trust relationships between military and civil authorities to increase efficiency and timeliness of information processing and exchange. PAUSE also makes use of AI and automation to extract, clarify, identify, categorize, locate, assess and most importantly fuse information from eye-witness sources (with variable trustworthiness) to improve the accuracy and accountability of decision-makers.

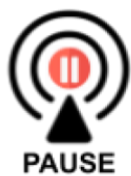

# Protected Assurance Understanding Situation & Entities (PAUSE)

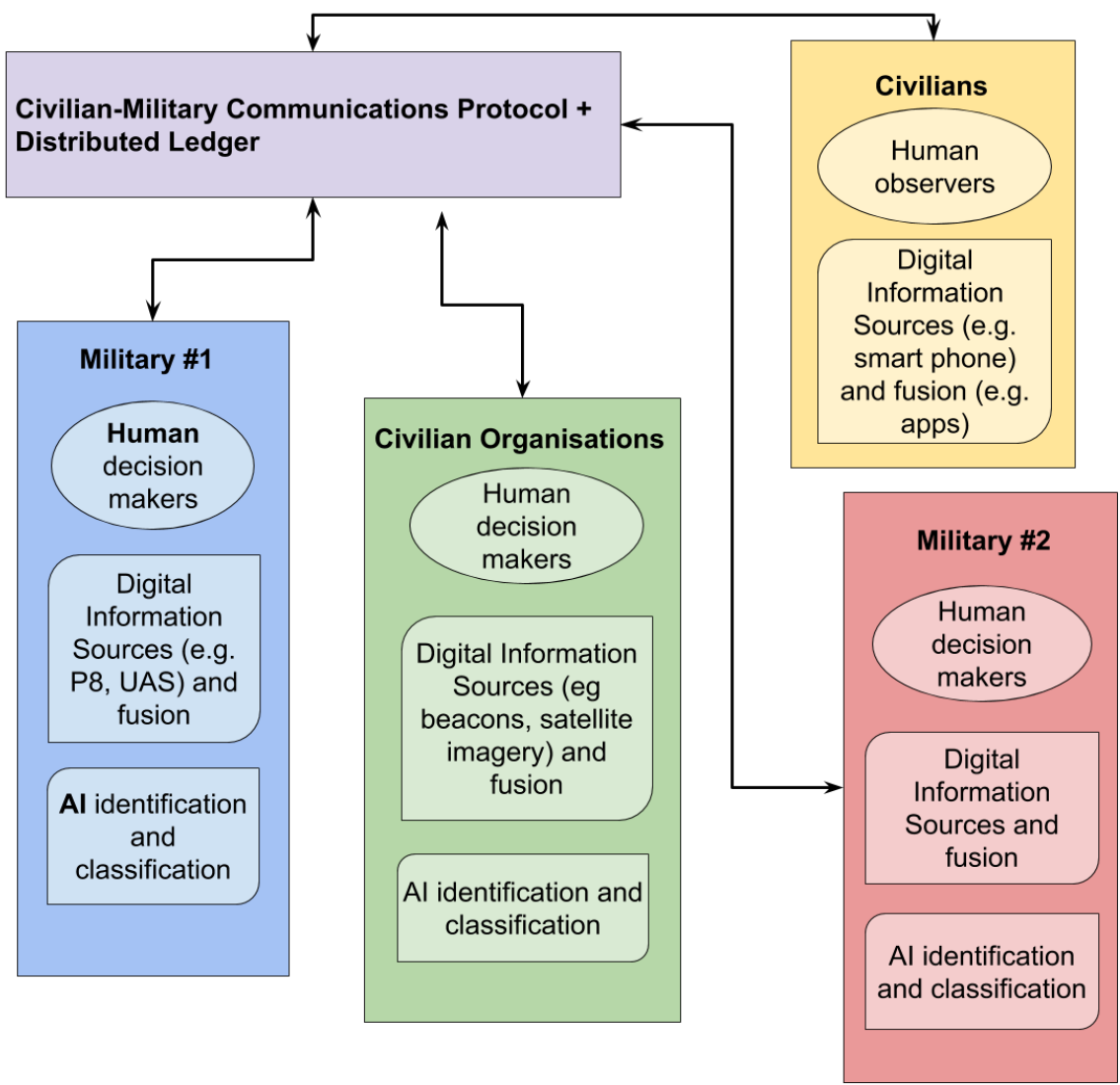

Figure 4. The PAUSE network relies on military and civilian organisations to bear responsibility for communication to the ledger including any human, digital and AI based information generated or used by these organisations. Civilians can broadcast including signs of surrender, proof of life and so forth. Civilian and military organisations could form their own judgments with regards to how to represent and respond to communications. In particular, civilian safety considerations must be managed if their messages put themselves at risk of harm by parties to a conflict. It might be agreed that an anonymity layer is required for civilian reports such as differential privacy (Dwork & Roth, 2014).

The PAUSE network recognizes that the locus of moral responsibility for any decision rests with humans, individually or within an organisation. Digital or AI assets support human decision-making. Humans must design AI and autonomous features within the context of human values using a human-centred approach (van den Hoven, 2013).

An AI classifier used by the military belongs to a State and the State must take end-to-end responsibility for their AI, noting that Protective AI and digital resources might be open-sourced. Many organisations, governments and militaries are actively progressing ethical frameworks to assist decision-makers in creating procedures and protocols for the development, testing, deployment, evaluation and adaptation of AI (International Committee of the Red Cross, 2019; Lopez, 2020). However, how to operationalize these frameworks within technologies is less advanced. In the first instance organisations must understand an AI's training data, inputs, functions, outputs, and boundaries (Robbins, 2019). Then organisations must situate the AI within a network of information where its limits and affordances are appropriately harnessed and restrained.

Protective AI and a civil-military communications protocol need to be "surfaced" in software applications in order for each organisation to use in accordance with their objectives and values. Technologies may be integrated within existing military and civilian software systems, or through new applications. For example, militaries seeking to abide by IHL are likely to want to layer data emerging from compliant communications with command and control data including military objectives and (ISR) data (Paul, Clarke, Triezenberg, Manheim, & Wilson, 2018). To better understand applications, we examine several case studies.

## 4.1 Case 1: Mapping a conflict zone in real time

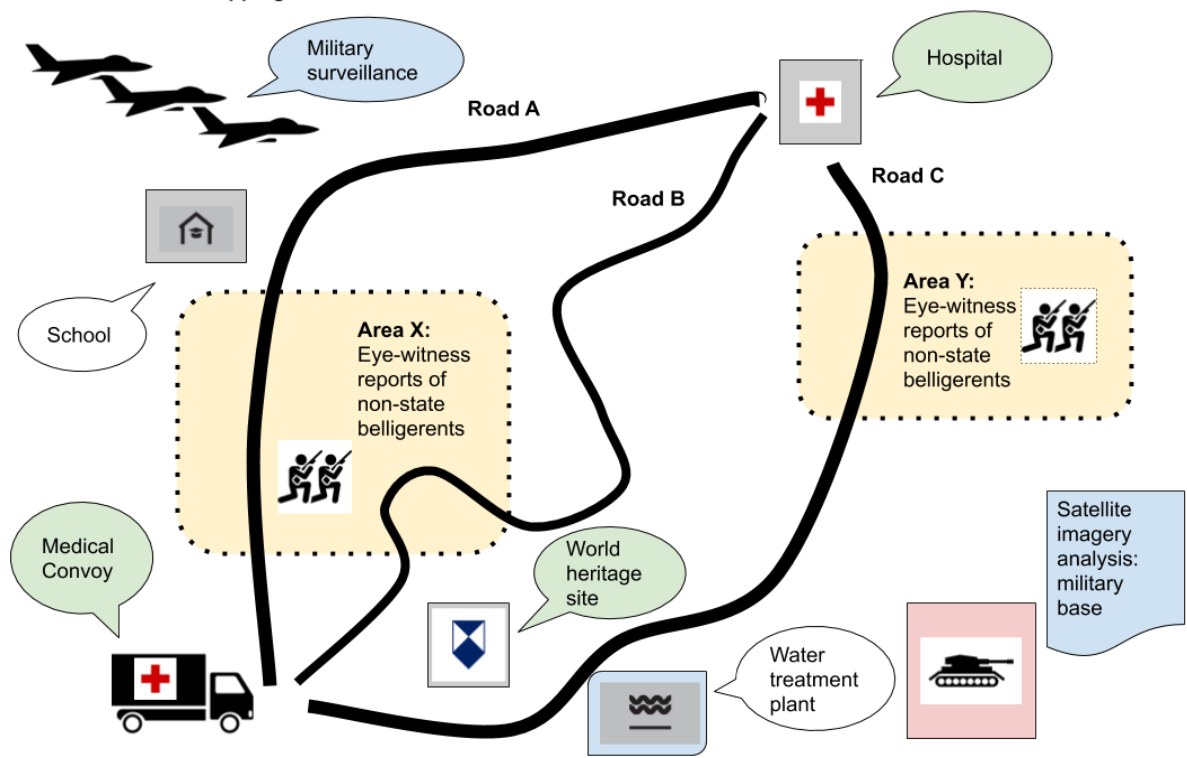

Figure 5. Example mapping of a conflict zone from a military perspective showing some protected objects and critical infrastructure, Whiteflag-compliant communications received from objects (white and green speech bubbles) and anonymised human observation reports (dotted boxes), combined with military analyses (blue reports)

The PAUSE system case in Figure 5, aims to "map" the location of protected sites and critical infrastructure in a conflict zone with real-time fully-traceable updates to reduce uncertainty, ambiguity and error. In a conflict zone there will be transmitters and receivers of information. Transmitters include protected object beacons, protected object detection sources (from AI), computer logs, video footage, satellite imagery, audio recordings. Humans in the zone may be both transmitters and receivers identifying people in need of assistance and/or surrendering (see Box 2: Digital and Trusted Surrender). Civilian communications may need special protections such as being anonymised using techniques such as differential privacy(Dwork & Roth, 2014). Assets that are responsibility of local government, NGOs, ICRC, UN etc. may also be transmitters and receivers. Using PAUSE each organisation decides what kinds of information to incorporate. Each nation or organisation would also create validation procedures for receivers, giving weight to information in relation to values, priorities and trust metrics. Transmitted data from some entities (e.g. ICRC) would be considered a trusted source.

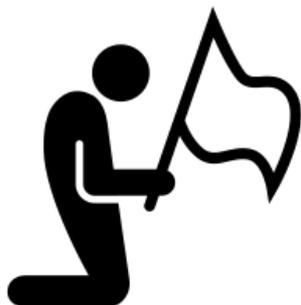

**Box 2: Digital and Trusted Surrender**

Humans could use the PAUSE system to register and better facilitate surrender and there is the potential for radio or other beacons to signal surrender. These signals would be visible to Combat systems and may require new laws to prevent unwanted exploitation. To illustrate, if the PAUSE system were available to US troops in the Persian Gulf War, (United States, 1992), Iraqi combatant lives may have been saved as information of surrender could be transmitted via a neutral communications channel like Whiteflag from beacons and/or text messages.

## 4.2  Case 2: Preparation to avoid incidents

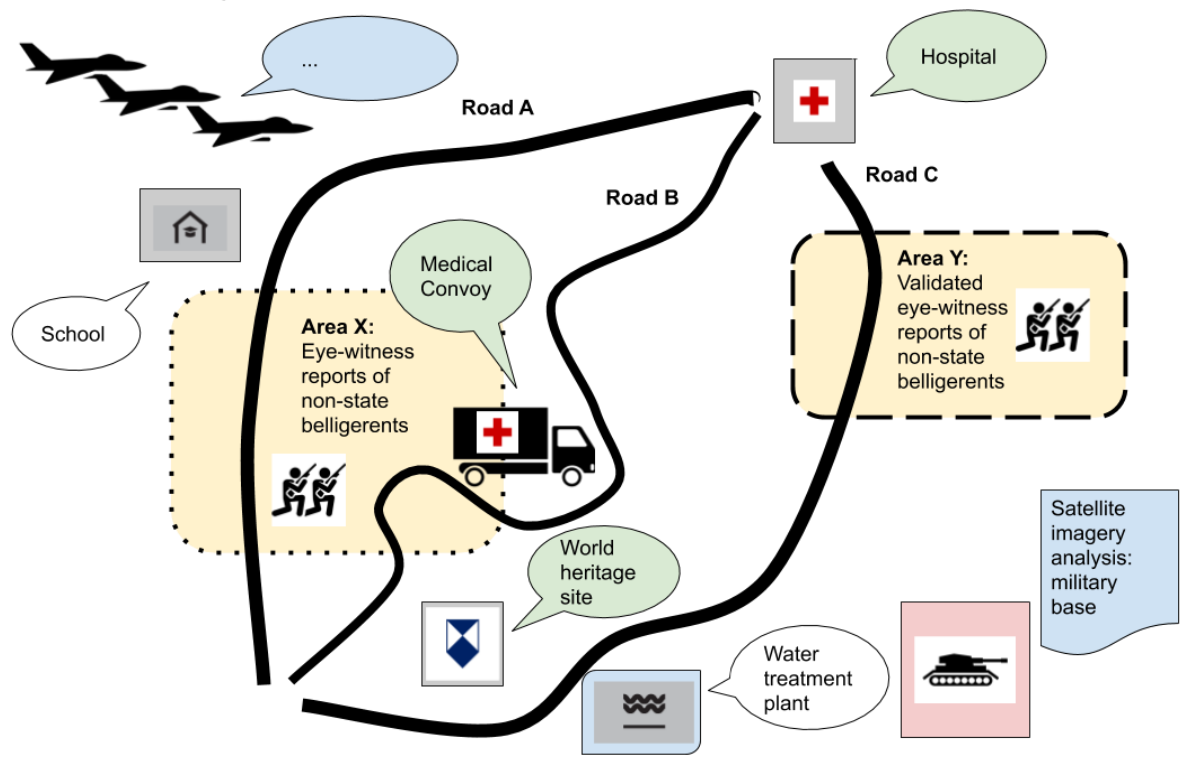

Figure 6. A humanitarian organisation using PAUSE to evaluate the risks to a resupply mission for humanitarian aid to a hospital via Road A, B or C. Road B could be chosen to minimise the likelihood of contact with: military activities such as the aircraft and non-state belligerents in Area X near Roads A and B, or a military base and non-state belligerents in Area Y near Road C. Analysis of anonymised human observations may lend more credibility to the hypothesis that non-state belligerents are near road C and less likely on Road A.

Best practice would be for States and organisations to incorporate PAUSE into their decision-making to avoid violations of IHL. Actors are expected to act in good faith to avoid an accident. Not using information provided via PAUSE may need justification if it becomes the de facto "standard". Figure 6 illustrates humanitarian organisations use of PAUSE to de-risk plans. In this scenario, PAUSE helps track the changing locations of protected objects such as the movement of medical supplies along Road B. Each organisation in a conflict should reassess any actions in light of this updated information.

A particular challenge for civilian and military organisations is to come to terms with how to respond to different information sources. This is why we consider models of information fusion. Once a normative model for decision-making by organisations is adopted, then agreement from diverse sources might be trusted more than the aggregation of views from a set of more homogeneous individuals (e.g. highly connected within social networks) (Schmidt et al., 2017). Information from trusted allies is likely to be weighted more highly than unfamiliar sources. Highly uncertain or contentious information might trigger ISR actions depending on the perceived risk of inaction. Thresholds for actions under different densities of evidence and under various levels of uncertainty would be decided within each State depending on their risk appetite and political will. Thus, the PAUSE network can be used to undertake a risk assessment of actions by both civilian and military groups.

## 4.3  Case 3: Dealing with Misinformation

The PAUSE network would enable organisations to establish appropriate levels of trust in sources and fusion of information with varying levels of trust and changing trust over time. This means that mistaken information or deliberate disinformation campaigns are anticipated and more readily identified and managed. Figure 7 illustrates a case where PAUSE uncovers military attempts to manipulate information and progress misinformation or disinformation campaigns.

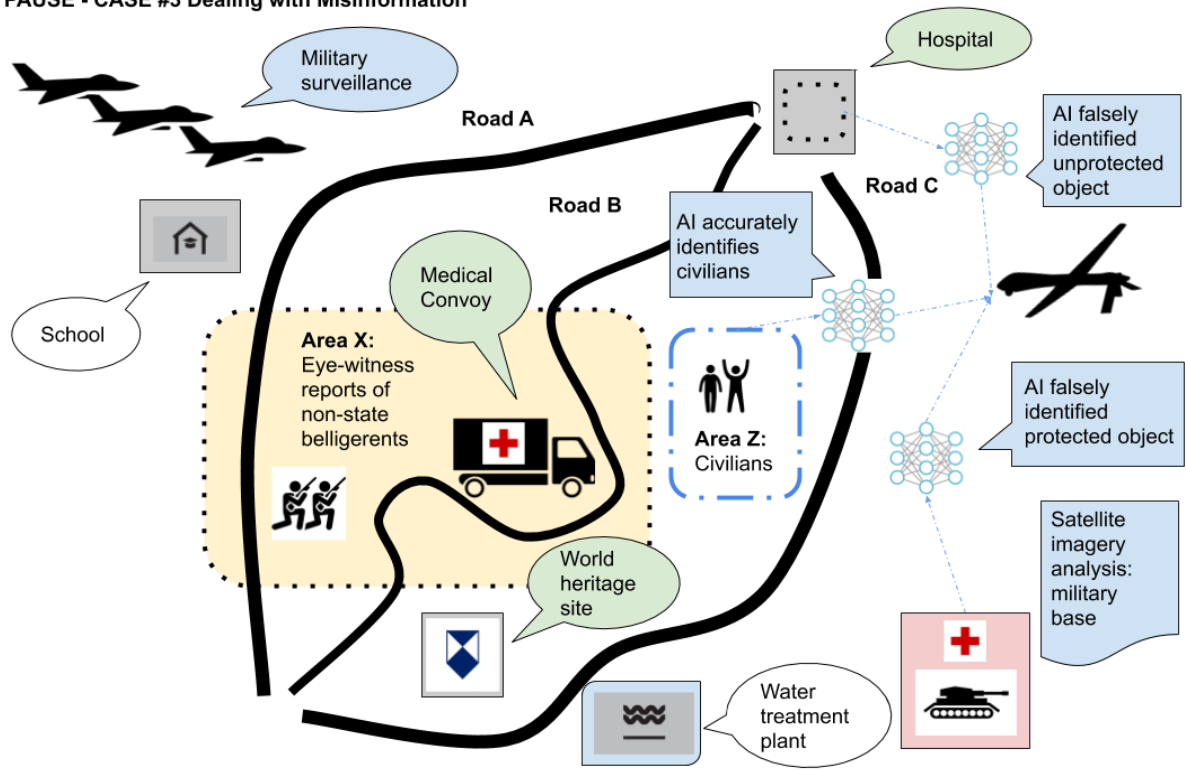

Figure 7. Belligerents try to unlawfully spoof ISR-equipped drones into believing military targets are protected (see red square) or that protected objects such as hospitals are neutral or military targets. The PAUSE system provides countervailing evidence so that disinformation campaigns can be questioned and mistaken information corrected. In this way, AI is part of the information system with many checks and balances.

Take for example a deliberate campaign to make protected symbols on protected objects unreadable to an AI visual feed or infrared sensor. If a military depended upon the AI processed inputs via drone feed to make decisions, then it could be fooled into believing a hospital was a legitimate military target. However, an AI within the PAUSE network would be augmented with counter veiling evidence that a hospital is at the stated location, moving decision-makers from fast intuitive Type I decisions to a more deliberate and reflective Type II thought-process (Kahneman, 2011) when appropriate.

An alternate scenario is that military objects would attempt to spoof AI classifiers into believing they were protected targets by painting protected icons on their sides, such as a tank painting a red cross. In the first instance, a smart AI would not just rely on identifying a red cross, but also the silhouette of a tank so that it would be cognisant of conflicting classifications. Additionally, if the AI-generated results were within the PAUSE network, alternate evidence would suggest both a) the absence of a known humanitarian vehicle in that area, e.g. the vehicle lacks an official ICRC beacon or similar and b) the known presence of military forces in that location, e.g. Satellite imagery reveals the massing of military hardware where vehicle is located. Militaries are likely to invest heavily in AI-design and complimentary ISR capabilities to autonomously sort through conflicted data to provide operators with timely advice on how to proceed.

*In sum,* there are three responses to the disinformation risk, 1) a range of AI techniques exist to deal with these situations with a high degree of confidence and 2) the PAUSE system, by allowing organisations and nations to set their own information fusion methods, would provide data-driven techniques to address adversarial uses of AI technically. The distributed ledger of objects and behaviours within conflict zones offer a reliable means to capture evidence to support new laws to make spoofing AI illegal. 3) a civilian-military protocol over a distributed ledger allows cross-checking with other trusted non-AI sources

# 5   Real World Case Study

The PAUSE conceptual framework may seem optimistic—indeed when first conceptualised by the authors in 2019, the political will to confront the complex realities that lead to civilian harm was not realised. However, the political climate has changed with the military withdrawal from Afghanistan by the United States in 2021. Specifically once the unlawful targeting of a civilian aid worker and death of civilians including children in Kabul 29 August revealed systematic inadequacies of the information systems used to determine applications of force by the United States (Lewis, 2021). Public awareness of US targeting (and mistargeting) has been heightened by news media campaigns, such as the New York Times *Civilian Casualty Files* (Department of Defense, 2022; Khan et al., 2022). Perhaps more importantly, USAID is now leading development of a new communication system using blockchain to "provide a secure means to improve the fidelity and completeness of the civilian environment picture" (see Box 2). The potential of PAUSE is being explored with political will and technical manifestation.

---

**Box 2**. Blockchain for Mitigating Harm to Civilians

Military targeting processes and supporting systems tend to rely on intelligence regarding the threat. While today's conflicts typically involve operations where civilians are present, the civilian environment is a blind spot for targeting. Information about civilian populations, humanitarian organizations, and civilian infrastructure is often inaccurate or incomplete. In our analysis, most incidents of civilian harm include deficiencies in knowledge of the civilian

---

environment: unnoticed civilians in proximity to the target, civilians misidentified as combatants, or humanitarian activities not recognized.

Fixing this blind spot by improving knowledge of the civilian environment would strengthen mitigation of civilian harm. To this end, USAID is leading the development of a new communication system—the Human Security Information System (HSIS)—to provide a secure means to improve the fidelity and completeness of the civilian environment picture.

HSIS will use blockchain technology, which creates a cryptographically secure set of records that is built up over time through distributed transactions. The use of blockchain in HSIS provides the security that only intended parties can access the information. It also features incorruptible records, creating an audit trail that can be leveraged for learning and accountability. The system will make it easier for organizations to report and update their information individually to increase timeliness of submissions, and in formats that promote accuracy. Civilian information such as critical infrastructure, cultural heritage sites, medical facilities, and other civilian objects can be reported either by individual organizations or collectively by a trusted agent on their behalf.

The standard format and data structure of HSIS promotes interoperability and integration with military systems, where the civilian information can be imported directly into military systems over data links like Link 16. HSIS can improve the overall quality and consistency of civilian information over current ad hoc processes, simplifying the work of militaries to integrate it into military systems and processes and creating a stronger foundation for effective mitigation of civilian harm.

Lewis, L. (2023). *Emerging technologies and civilian harm mitigation*. Center for Naval Analysis.

# 6 Conclusion

Although the "fog of war" limiting situation awareness may never entirely clear, it should not be held as the standard to explain-away civilian targets, justify collateral damage, or give up on the pursuit of better technologies. In a world increasingly plagued by extreme events yet brimming with information, more proactive responses are needed. We have outlined our suggestion that a trusted human-AI network requires diverse information sources to be appropriately fused and a communications protocol based on blockchains for civilian-military information adopted. We have illustrated the potential for AI to process humanitarian and critical infrastructure information, but acknowledge the risks of dis- and misinformation that must be managed through individual, organisational and inter-organisational information optimisation. A trusted network based on these technologies holds the potential to improve the efficiency and timeliness of humanitarian actions in conflict and disaster settings. In this way, States, non-state actors, humanitarian organisations and NGOs might change the technical means by which they communicate in conflict without threatening their neutrality or security, while making reliable use of eye-witness materials and social media.

**Acknowledgements:** The authors would like to acknowledge the contributions of A/Prof Simon Ng, Dr Beth Cardier and Dr Bianca Baggiarini in the preparation of this manuscript

The research for this paper received funding from the Australian Government through Trusted Autonomous Systems, a Defence Cooperative Research Centre funded through the Next Generation Technologies Fund.

# 8  Appendices

## 8.1  Appendix A. Whiteflag protocol functional message categories and examples

| FUNCTION | DESCRIPTION | EXAMPLES |
|---|---|---|
| **PROTECTIVE SIGNS** | Signs to mark objects under the protection of international law | Hospitals, Safety Zones, White Flag, Humanitarian Convoys, Cultural Property, Medical Units |
| **EMERGENCY SIGNALS** | Signals to send an emergency signal when in need of assistance | Emergency Beacon, Distress Signal |
| **DANGER SIGNS** | Signs to mark a location or area of imminent danger | Mark danger such as an area under attack, land mines, disasters, etc. |
| **STATUS SIGNALS** | Signals to provide the status of an object, or specifically for persons: give a proof of life | Personal beacon on individuals or confirmation of persons for assistance, status of critical infrastructure. |
| **INFRASTRUCTURE SIGNS** | Signs to mark critical infrastructure | Roads, schools, utilities, water treatment, hospitals, power plants etc. |
| **MISSION SIGNALS** | Signals to provide information on activities undertaken during a mission | Intentions of objects such as convoys as they progress and adapt through a mission, deconfliction with military operations |
| **REQUEST SIGNALS** | Signals to perform requests to other parties | Requests for area access, cease fire, etc. |
| **RESOURCE MESSAGES** | Messages to point to an internet resource with additional information | Additional information from official websites on cultural property, minefields, news feeds, authenticated photographs, etc. |
| **FREE TEXT MESSAGES** | Messages to send a free text string to clarify and provide context to other messages | Supplementary commentary to enable further clarification of ambiguous events |

Whiteflag messages sent on a blockchain are pre-defined and based on international rules and standards for armed conflicts and disasters. This ensures interoperability between any Whiteflag-capable system and a common understanding between the communicating parties. The messages should be seen as digital equivalents of physical signs and communication signals marking entities protected under international humanitarian law, critical infrastructures, emergencies, and danger zones such as minefields etc. These messages enable entities protected under humanitarian law to make themselves known in real-time to parties they trust for deconfliction; disclose real-time critical information to trusted parties to improve overall shared situational awareness; and allow organisations to notify others of their planned and ongoing activities. For natural disasters on the other hand, close collaboration between aid workers, affected people and the general public is important. In those circumstances, Whiteflag may be used openly with information made available by anybody to everyone to quickly create and disseminate near real-time information about imminent dangers, available aid, etc. to create shared situational awareness. ("The Whiteflag Protocol," 2020).

## 8.2  Appendix B: Decision error considerations for MinAI-enabled weapon systems

| Truth | Operator Perceived | Machine Perceived | State | Consequences |
|---|---|---|---|---|
| Protected | Protected | Protected | Correct protection | Protection achieved |
| Protected | Protected | Not Protected | Correct protection | Protection achieved as human prohibits machine to engage. |
| Protected | Not protected | Protected | Correct protection | Protection achieved as machine prohibits engagement |
| Protected | Not protected | Not protected | Protection fail | Protection failure |
| Not protected | Protected | Protected | False protection | Military objective not achieved |
| Not protected | Protected | Not protected | False protection | Military objective not achieved as human prohibits machine to engage |
| Not protected | Not protected | Protected | False protection | Military objective not achieved as machine prohibits engagement |
| Not protected | Not protected | Not protected | Unprotected | Military objective achieved within IHL/ILAC-boundaries |

.